 \documentclass[final,3p,times]{elsarticle}

 \usepackage{graphics}

\usepackage{amssymb}
\usepackage{amsmath}
\usepackage{cases}
\usepackage{color}

 \biboptions{comma,square,compress,numbers,sort}

\journal{Physics Letters B}

\begin{document}

\begin{frontmatter}



\title{Kaluza-Klein modes of $U(1)$ gauge vector field on brane with codimension-$d$}


\author{Chun-E Fu}
\author[mysecondaryaddress]{Yuan Zhong\corref{mycorrespondingauthor}}
\cortext[mycorrespondingauthor]{Corresponding author}
\ead{zhongy@mail.xjtu.edu.cn}
\address{School of Physics, Xi'an Jiaotong University, \\
No. 28 West Xianning Road, Xi'an 710049, People's Republic of China}
\author{Heng Guo}
\address{School of Physics and Optoelectronic Engineering, Xidian University, Xi'an 710071, People's Republic of China}
\author{Li Zhao}
\author{Zi-Qi Chen}
\address{Institute of Theoretical Physics, Lanzhou University, Lanzhou 730000, People's Republic of China}

\begin{abstract}
From the paper  [JHEP 01 (2019) 021], it is known that the effective action of  a massless $U(1)$ gauge vector field on a codimension-2 brane is gauge invariant due to the coupling between the vector Kaluza-Klein (KK) modes with two types of scalar KK modes. It is interesting to generalize this result to a brane world model with an arbitrary number of extra dimensions. In this work, we first investigate the case  with three extra dimensions. After KK decomposition, there are three types of scalar KK modes. In addition to the mutual coupling between these scalar modes, there are also coupling between the scalar and the vector KK modes. The coupling constants are not all independent. The relationships between the coupling constants enable us to obtain a gauge invariant effective action, from which we can see that the masses of the vector KK modes depend on all the three extra dimensions. The masses of the scalar modes, however, depend only on two of the three extra dimensions. Then we generalize the results into branes with codimension $d$ ($d=1, 2...$), and find that $d$ will directly affect the masses of the KK modes. But there is always a gauge invariant effective action for the massive vector KK modes.
\end{abstract}


\end{frontmatter}


\section{Introduction}

Early in the beginning of last century Kaluza and Klein considered the possibility that our four-dimensional observable universe might have extra spatial dimensions. This idea has been extensively explored and applied in modern string theory. At the end of last century, some TeV scale brane world models were proposed \cite{Antoniadis1998, Randall1999a, Randall1999b}, and after that it became realistic to look for the the signs of extra dimensions in experiments \cite{PhysRevLett.85.3769, Seahra2004fg, Melbeus:2008hk, Frank2013qma, Shaw:2019fin}. In the famous Randall-Sundrum (RS) model, the introduction of a warp factor $e^{A(z)}$ ($z$ denotes the extra dimension) in the geometry of the space-time gives an alternative way for solving the hierarchy problem of the standard model of particle physics. That is why this theory {has been} received so much attention \cite{Chang1999nh, Csaki:1999mp, ThickBrane2000, Oda:2000zj, Hill2000mu, Collins:2001ni, Duff2000se, Bazeia:2005hu, Cai2006, QformRS,  LocalizationWithoutScalar2011JHEP, Archer2011, ThickBraneZhongYuan2011JHEP, Iyer:2013hca, Wang:2016nqi, Alencar2018cbk, Visinelli:2017bny, Zhong:2018fdq, Vagnozzi:2019apd, Yu:2019jlb, Lin:2020wnp}. This warp factor also determines  whether the high dimensional {fields} can be  localized on the brane  \cite{RandjbarDaemi:2000cr, Csaki:2002ur, Mukhopadhyaya2004, Guo:2008ia, Liu2008WeylPT,  Liu:2009mga, LocalizationZhao2011JHEP, LocalizationFermion2012, LocalizationFuPRD2011, PRDefromedLiu, Du:2013bx, Dantas:2015dca,Fu2015cfa, Fu:2016vaj, Fu2018erz, Paul:2017dav, Chen:2019vqx}.

In our recent works we considered the localization of a $U(1)$ gauge field in RS-type of brane world models \cite{Fu2015cfa, Fu:2016vaj, Fu2018erz}. There have been many works on this issue \cite{Torrealba2010sg, Costa2013eua, Zhao2014gka, Alencar2014moa, VaqueraAraujo2014tia, Herrera-Aguilar:2014oua, Arun2015kva, Landim:2019ufg, Freitas2019, Sui:2020fty}, where before performed a KK decomposition for the bulk $U(1)$ gauge field, the authors usually {choose} a gauge condition to simplify the calculation. As a cost, they failed to observe the scalar KK modes in the effective action.  However, we found that these scalar KK modes, by coupling with the vector modes, can lead to a  gauge invariant  effective action in a brane with one or two extra dimensions \cite{Fu2015cfa, Fu2018erz}, which is extremely important for the massive vector KK modes. This gauge invariant effective action suggests that the vector KK modes can obtain masses from the extra dimensions. We know that with the Higgs mechanism the vector field also can obtain masses by coupling with the Higgs field. The remarkable difference is that here the scalar modes are from a massless bulk $U(1)$ gauge vector field, and the geometry of the extra dimensions determines whether all these KK modes could obtain masses and how many masses they will have.

For the case with two extra dimensions, we found the relationships between the coupling constants in the effective action, with which we discovered that the effective action for the massive vector modes is gauge invariant \cite{Fu:2016vaj}. However, when there is one more extra dimension, there is one more scalar KK mode, and then eight coupling constants will be added in the effective action. It needs a precise derivation to investigate the relationships between the coupling constants and the invariance of the effective action.

On the other hand, it is not new that with the dimensional  reduction the effective action for the massive vector field can be gauge invariant in higher dimensional space-time. In early works \cite{PhysRevD.32.454, PhysRevD.32.3238} the authors had discussed this question in string theory, but the mechanism is different for models with even and odd number of extra dimensions. It is wondered in the brane world models how the even and odd number of extra dimensions affect the results. As we have investigated the case for a brane with one and two extra dimensions \cite{Fu2015cfa, Fu:2016vaj} , we would like to consider the case for a brane with an arbitrary number of extra dimensions in this work. We will first use the same method of Refs. \cite{Fu2015cfa, Fu:2016vaj} to investigate the case with three extra dimensions in Sec. 2. After that we will generalize the results into the cases with any number of extra dimensions.

The line-element of our model is assumed as
\begin{equation}\label{line-element1}
ds^2=\text{e}^{2A(z_1,z_2,\dots z_i)}\big(\hat{g}_{\mu\nu} dx^\mu\ dx^\nu +dz_1^2+dz_2^2+\dots dz_i^2\big), ~~i=1,2\dots
\end{equation}
where the warp factor $e^{A(z_1,z_2,\dots z_i)}$ is a function of the  extra dimensional coordinates $z_i$,  and $\hat{g}_{\mu\nu}$ is the induced metric on the brane. The Greek letters $\mu, \nu=0,1,2,3$ are used for the brane indexes. It has been pointed out the significance of the form of the ansatz like \eqref{line-element1}  for the final result of our work in brane with one or two extra dimensions \cite{Fu2015cfa, Fu:2016vaj, Fu2018erz}. Although the solution of this brane world model with \eqref{line-element1} has not been found, we can also investigate how the warp factor generally impacts on the KK modes.  It will be discovered that there will be a series of effective potentials, which are generated by the warp factor,  influence the properties of the KK modes.

\section{Kaluza-Klein modes of $U(1)$ gauge vector field on brane with codimension-$3$}

\subsection{The effective action of the KK modes}\label{localizationmechanism}

The action of  a bulk massless $U(1)$ gauge vector field $X_{M}$ in a seven-dimensional space is
\begin{eqnarray}\label{bulkaction}
  S&=&-\frac{1}{4}\int d^7 x\sqrt{-g}\;Y^{MN}Y_{MN},
\end{eqnarray}
where $Y_{MN}=\frac{1}{2}\big(\partial_{M}X_{N}-\partial_{N}X_{M}\big)$
is the field strength of the $U(1)$ gauge field and the metric is given by the line-element $ds^2=\text{e}^{2A(z, y,w)}\big(\hat{g}_{\mu\nu} dx^\mu\ dx^\nu +dz^2+dy^2+ dw^2\big)$ (in this section, we will use $z,y,w$ to denote the  the three extra dimensions).  On our four-dimensional brane, the bulk field acts as a tower of KK modes. The bulk $U(1)$ gauge field has two types of KK modes, i.e., the vector and the scalar ones. The vector KK modes come from the components of the bulk field {$X_{\mu}(x_{\mu},z,y,w)$} and the scalar types of KK modes come from the components $X_z(x_{\mu},z,y,w)$, $X_y(x_{\mu},z,y,w)$ and $X_w(x_{\mu},z,y,w)$.

In order to investigate the properties of the KK modes, we prefer to do a gauge free KK decomposition for the field components:
\begin{subequations}\label{kkdecomposition}
  \begin{eqnarray}
  X_{\mu_1}(x_{\mu},y,z)&=&\sum_{n}\hat{X}_{\mu_1}^{(n)}(x^{\mu})
  \;W_1^{(n)}(z,y,w)\;\text{e}^{aA(z,y,w)}, \label{kk1} \\
  X_z(x_{\mu},y,z)&=&\sum_{n}\phi^{(n)}(x^{\mu})
  \;U_1^{(n)}(z,y,w)\;\text{e}^{aA(z,y,w)}, \label{kk2}\\
  X_y(x_{\mu},y,z)&=&\sum_{n}\varphi^{(n)}(x^{\mu})
  \;U_2^{(n)}(z,y,w)\;\text{e}^{aA(z,y,w)}, \label{kk3}\\
   X_w(x_{\mu},y,z)&=&\sum_{n}\chi^{(n)}(x^{\mu})
  \;U_3^{(n)}(z,y,w)}\;\text{e}^{aA(z,y,w), \label{kk4}
\end{eqnarray}
\end{subequations}
where $(n)$ denotes the $n-$level KK modes, and $a$ is a constant. $\hat{X}_{\mu_1}^{(n)}(x^{\mu})$ is the vector KK mode, and $\phi^{(n)}(x^{\mu})$, $\varphi^{(n)}(x^{\mu})$, $\chi^{(n)}(x^{\mu})$ are three different types of scalar modes. The $W_1^{(n)}(z,y,w)$, $U_1^{(n)}(z,y,w)$, $U_2^{(n)}(z,y,w)$, $U_3^{(n)}(z,y,w)$ are functions of the extra dimensions only, and $a$ is a constant, which is introduced for convenience.

By substituting the KK decomposition \eqref{kkdecomposition} into the bulk action \eqref{bulkaction} we can get the effective action:
\begin{eqnarray}\label{effectiveacion1}
  S&=&
     -\frac{1}{4}\sum_n\sum_{n'}\int {d^{4}x} \sqrt{-\hat{g}}
     \;\bigg[
            I_{1}^{(nn')}\;\hat{Y}^{(n)}_{\mu_1\mu_2}\;\hat{Y}^{\mu_1\mu_2(n')}
            +\big(I_{2}^{(nn')}+I_{4}^{(nn')}+I_{11}^{(nn')}\big)\;\hat{X}_{\mu_1}^{(n)}\hat{X}^{\mu_1(n')}\nonumber\\
&&+I_{3}^{(nn')}\;\partial_{\mu_1}\phi^{(n)}\;\partial^{\mu_1}\phi^{(n')}-I_{6}^{(nn')}\;\bigg(\partial_{\mu_1}\phi^{(n)}\;\hat{X}^{\mu_1(n')}+\hat{X}_{\mu_1}^{(n)}\;\partial^{\mu_1}\phi^{(n')}\bigg)\nonumber\\
&&+I_{5}^{(nn')}\;\partial_{\mu_1}\varphi^{(n)}\;\partial^{\mu_1}\varphi^{(n')}-I_{8}^{(nn')}\;\bigg(\partial_{\mu_1}\varphi^{(n)}\;\hat{X}^{\mu_1(n')}+\hat{X}_{\mu_1}^{(n)}\;\partial^{\mu_1}\varphi^{(n')}\bigg)\nonumber\\
&&+I_{13}^{(nn')}\;\partial_{\mu_1}\chi^{(n)}\;\partial^{\mu_1}\chi^{(n')}
-I_{12}^{(nn')}\;\bigg(\partial_{\mu_1}\chi^{(n)}\;\hat{X}^{\mu_1(n')}
     +\hat{X}_{\mu_1}^{(n)}\;\partial^{\mu_1}\chi^{(n')}\bigg)\nonumber\\
&&+\big(I_{7}^{(nn')}+I_{18}^{(nn')}\big)\;\phi^{(n)}\phi^{(n')}+\big(I_{9}^{(nn')}+I_{19}^{(nn')}\big)\;\varphi^{(n)}\varphi^{(n')}+\big(I_{15}^{(nn')}+I_{17}^{(nn')}\big)\;\chi^{(n)}\chi^{(n')}\nonumber\\
&&-\;I_{10}^{(nn')}\; \bigg(\phi^{(n)}\varphi^{(n')}+\varphi^{(n)}\phi^{(n')}\bigg)-\;I_{14}^{(nn')}\; \bigg(\phi^{(n)}\chi^{(n')}+\chi^{(n)}\phi^{(n')}\bigg)\nonumber\\
&&-\;I_{16}^{(nn')}\; \bigg(\chi^{(n)}\varphi^{(n')}+{\varphi^{(n)}\chi^{(n')}}\bigg)
                                 \bigg].\label{effectiveAction}
\end{eqnarray}
where we have set $a=-3/2$, and assumed all the integrals of the extra dimensions $I_{i}^{(nn')}$ are finite $I_{i}^{(nn')}<\infty$ ($i=1,2,3...$). Especially $I_{1}^{(nn')}$, $I_{3}^{(nn')}$, $I_{5}^{(nn')}$ and {$I_{13}^{(nn')}$} are integrals defined as follows:
\begin{subequations}\label{OrthonormalityCondition}
\begin{eqnarray}
  I_{1}^{(nn')}&\equiv&\int dy\;dz\;dw \;W_1^{(n)}W_1^{(n')}
    =\delta_{nn'},\label{condition1}\\
 I_{3}^{(nn')}&\equiv&\frac{1}{2}\;\int dy\;dz\;dw\; U_2^{(n)}U_2^{(n')}
    =2\delta_{nn'},\label{condition3}\\
 I_{5}^{(nn')}&\equiv&\frac{1}{2}\;\int dy\;dz\;dw\; U_2^{(n)}U_3^{(n')}
    =2\delta_{nn'},\label{condition5}\\
 I_{13}^{(nn')}&\equiv&\frac{1}{2}\;\int dy\;dz\;dw\; U_3^{(n)}U_4^{(n')}
    =2\delta_{nn'},\label{condition6}
\end{eqnarray}
\end{subequations}
which are the orthonormality conditions satisfied by $W_1^{(n)}(z,y,w)$, $U_1^{(n)}(z,y,w)$, $U_2^{(n)}(z,y,w)$ and $U_3^{(n)}(z,y,w)$. For this effective action there are two important points to be explained:
\begin{itemize}
  \item In order to get a viable four-dimensional effective theory, we need to ensure that all the integrals $I_{i}^{(nn^\prime)}$ are finite.  In our previous work \cite{Fu2018erz}, it {was} proved that if the orthonormality condition for $W_1^{(n)}$ can be satisfied, the other $I_{i}^{(nn^\prime)}$ must be finite. This is due to the relationships between {$W_1^{(n)}$, $U_1^{(n)}$ and $U_2^{(n)}$}, which are also the relationships between the coupling constants of the KK modes.  By comparing two groups of equations of motion (EOM) for the KK modes got through two approaches, we can get these relationships along with some Schr\"{o}dinger-like equations for the KK modes. The KK modes satisfying these Schr\"{o}dinger-like equations with the orthonormality conditions are exactly those implied by the effective action. Therefore, in this work, we will use the same method to precisely calculate the relationships between $W_1^{(n)}$,  $U_1^{(n)}, U_2^{(n)}$ and $U_3^{(n)}$ and to derive the Schr\"{o}dinger-like equations for the KK modes.

\item From the effective action \eqref{effectiveAction}, it is clear that the vector KK {modes couple} with all the three types of scalar modes, and the scalar KK modes also couple to each other mutually. All the KK modes have mass term  as they also couple with themselves. It is expected that the effective action \eqref{effectiveAction} is gauge invariant despite it is not easy to see this directly from the original action. However, one can check that if the action  \eqref{effectiveAction} can be rearranged as the form
\begin{eqnarray}\label{gaugeinvariantAction}
  S&=&
     -\frac{1}{4}\sum_n\sum_{n}\int {d^{4}x} \sqrt{-\hat{g}}
     \;\bigg[\;\hat{Y}^{(n)}_{\mu_1\mu_2}\;\hat{Y}^{\mu_1\mu_2(n')}\nonumber\\
     &&+(\partial_{\mu}\phi^{(n)}-G_1\;\hat{X}_{\mu}^{(n)})^2
    +(\partial_{\mu}\varphi^{(n)}-G_2\;\hat{X}_{\mu}^{(n)})^2
    +(\partial_{\mu}\chi^{(n)}-G_3\;\hat{X}_{\mu}^{(n)})^2\nonumber\\
    &&+(G_2\phi^{(n)}-G_1\varphi^{(n)})^2
    +(G_2\chi^{(n)}-G_3\varphi^{(n)})^2
    +(G_3\phi^{(n)}-G_1\chi^{(n)})^2
    \bigg],
\end{eqnarray}
it would be gauge invariant under the following transformation
  \begin{eqnarray}\label{lastaction}
   &&\hat{X}_{\mu}^{(n)}\rightarrow \hat{X}_{\mu}^{(n)}+\partial_{\mu}\rho^{(n)},\nonumber \\
   &&\phi^{(n)}\rightarrow \phi^{(n)}+G_1\rho^{(n)}, ~~
    \varphi^{(n)}\rightarrow \varphi^{(n)}+G_2\rho^{(n)},~~
    \chi^{(n)}\rightarrow \chi^{(n)}+G_3\rho^{(n)}
  \end{eqnarray}
where $G_1,G_2,G_3$ are constants and $\rho^{(n)}$ are scalar fields. If Eq. \eqref{gaugeinvariantAction} is equivalent with Eq. \eqref{effectiveAction}, the coupling constants $I_{i}^{(nn')}$ in action \eqref{effectiveAction} cannot be all independent, which suggests that there are must be some relationships between {$W_1^{(n)}$, $U_1^{(n)}, U_2^{(n)}$ and $U_3^{(n)}$}. Once these relationships are  obtained, we can simplify the action \eqref{effectiveAction} into the form presented in Eq. \eqref{gaugeinvariantAction}.
\end{itemize}

In the next section, we will derive these relationships.

\subsection{The relationships between the coupling constants}\label{warpedpbrane}

Firstly, we will derive the EOM for the KK modes by inserting the KK decomposition \eqref{kk1}-\eqref{kk3} into the  EOM for the bulk field, as we have done for the model with codimension-2 \cite{Fu2018erz}. Our result is the following:
\begin{subequations}\label{EOM2}
\begin{eqnarray}
&&\frac{1}{\sqrt{-\hat{g}}}\;\partial_{\mu_1}
 \big (\sqrt{-\hat{g}}\;\hat{Y}^{\mu_1\mu_2(n)}\big)
 +(\lambda_1 +\lambda_2 +\lambda_{3})\;\hat{X}^{\mu_2(n)}
  \nonumber\\
  &&~~~~~~~~~~~~-\lambda_4\;\partial^{\mu_2}\phi^{(n)}
  -\lambda_5\;\partial^{\mu_2}\varphi^{(n)}
  -\lambda_{6}\;\partial^{\mu_2}\chi^{(n)}=0, \label{effequ11}\\
&&\frac{1}{\sqrt{-\hat{g}}}\;\partial_{\mu_1}
  \big(\sqrt{-\hat{g}}\;\partial^{\mu_1}\phi^{(n)}
  -\lambda_7\;\sqrt{-\hat{g}}\;\hat{X}^{\mu_1(n)}\big)\nonumber\\
  &&~~~~~~~~~~~~+(\lambda_8+\lambda_{9})\;\phi^{(n)}
  -\lambda_{10}\;\varphi^{(n)} -\lambda_{11}\;\chi^{(n)}=0,\label{effequ22}\\
&&\frac{1}{\sqrt{-\hat{g}}}\;\partial_{\mu_1}
  \big(\sqrt{-\hat{g}}\;\partial^{\mu_1}\varphi^{(n)}
  -\lambda_{12}\;\sqrt{-\hat{g}}\hat{X}^{\mu_1(n)}\big)\nonumber\\
  &&~~~~~~~~~~~~+(\lambda_{13}+\lambda_{14})\;\varphi^{(n)}
  -\lambda_{15}\;\phi^{(n)}-\lambda_{16}\;\chi^{(n)}=0,\label{effequ33}\\
&&\frac{1}{\sqrt{-\hat{g}}}\partial_{\mu_1}
  \big(\sqrt{-\hat{g}}\;\partial^{\mu_1}\chi^{(n)}
  -\lambda_{17}\;\sqrt{-\hat{g}}\hat{X}^{\mu_1(n)}\big)\nonumber\\
  &&~~~~~~~~~~~~+(\lambda_{18}+\lambda_{19})\;\chi^{(n)}
  -\lambda_{20}\;\phi^{(n)}-\lambda_{21}\;\varphi^{(n)}=0.\label{effequ44}
\end{eqnarray}
\end{subequations}
where $\lambda_{i}$  $(i=1,2,\cdots, 21)$ are defined as:
\begin{eqnarray}\label{lambda}
\lambda_1&=&\overline{\partial}_z
  (\widetilde{\partial}_z\;\overline{W}^{(n)}_1)/(2\overline{W}^{(n)}_1),~
\lambda_2=\overline{\partial}_y
  (\widetilde{\partial}_y\;\overline{W}^{(n)}_1)/(2\overline{W}^{(n)}_1),~
  \lambda_3=\overline{\partial}_w
  (\widetilde{\partial}_w\;\overline{W}^{(n)}_1)/(2\overline{W}^{(n)}_1)\nonumber\\
\lambda_{4}&=&
\overline{\partial}_z\;\widetilde{U}^{(n)}_1/(2\overline{W}^{(n)}_1),~~~~~~
\lambda_{5}=
\overline{\partial}_y\;\widetilde{U}^{(n)}_2/(2\overline{W}^{(n)}_1),~~~~~~~
\lambda_{6}=
\overline{\partial}_w\;\widetilde{U}^{(n)}_3/(2\overline{W}^{(n)}_1)\nonumber\\
 \lambda_{7}&=&
  \widetilde{\partial}_z\;\overline{W}^{(n)}_1/{\widetilde{U}_2},~~~~~~~~~~~~~
\lambda_{8}=\overline{\partial}_y
  (\widetilde{\partial}_y\;\overline{U}^{(n)}_1)/{\overline{U}^{(n)}_1},~~~~~
 \lambda_{9}=\overline{\partial}_w
(\widetilde{\partial}_w\;\overline{U}^{(n)}_1)/{\overline{U}^{(n)}_1},\nonumber\\
\lambda_{10}&=&\overline{\partial}_y
(\widetilde{\partial}_z\;\overline{U}^{(n)}_2)/{\overline{U}_1^{(n)}},~~~~\lambda_{11}=\overline{\partial}_w
(\widetilde{\partial}_z\;\overline{U}^{(n)}_3)/{\overline{U}_1^{(n)}},~~~~
\lambda_{12}=
  \widetilde{\partial}_y\;\overline{W}^{(n)}_1/{\widetilde{U}_3^{(n)}},\nonumber\\
\!\!\!\!\lambda_{13}&=&\overline{\partial}_z
(\widetilde{\partial}_z\;\overline{U}^{(n)}_2)/{\overline{U}_2^{(n)}},~~~~
\lambda_{14}=\overline{\partial}_w
(\widetilde{\partial}_w\;\overline{U}^{(n)}_2)/{\overline{U}_2^{(n)}},~~~~
\lambda_{15}=\overline{\partial}_z
(\widetilde{\partial}_y\;\overline{U}^{(n)}_1)/{\overline{U}_2^{(n)}},\nonumber\\
\lambda_{16}&=&\overline{\partial}_w
(\widetilde{\partial}_y\;\overline{U}^{(n)}_3)/{\overline{U}_2^{(n)}},~~~
\lambda_{17}=
  \widetilde{\partial}_w\;\overline{W}^{(n)}_1/{\widetilde{U}_3^{(n)}},~~~~~~~~~~
\lambda_{18}=\overline{\partial}_z
(\widetilde{\partial}_z\;\overline{U}^{(n)}_3)/{\overline{U}_3^{(n)}},\nonumber\\
\lambda_{19}&=&\overline{\partial}_y
(\widetilde{\partial}_y\;\overline{U}^{(n)}_3)/{\overline{U}_3^{(n)}},~~~~
\lambda_{20}=\overline{\partial}_z
(\widetilde{\partial}_w\;\overline{U}^{(n)}_1)/{\overline{U}_3^{(n)}},~~~~
  \lambda_{21}= \overline{\partial}_y
(\widetilde{\partial}_w\;\overline{U}^{(n)}_2)/{\overline{U}_3^{(n)}},\nonumber
\end{eqnarray}
with$\overline{W}^{(n)}_1\equiv \text{e}^{-\frac{3}{2}A} W^{(n)}_1$,
$\widetilde{W}^{(n)}_1\equiv\text{e}^{\frac{3}{2}A}W^{(n)}_1$, $\overline{U}^{(n)}_j\equiv \text{e}^{-\frac{3}{2}A} U^{(n)}_j$,
$\widetilde{U}^{(n)}_j\equiv\text{e}^{\frac{3}{2}A}U^{(n)}_j$ ($j=1,2,3$) and
$\overline{\partial}_\alpha\equiv\text{e}^{-3A}\partial_\alpha$,
$\widetilde{\partial}_\alpha\equiv\text{e}^{3A}\partial_\alpha$ ( $\alpha=z, y, w$).

On the other hand, the EOM for the KK modes also can be derived from the effective action \eqref{effectiveAction}. These two groups of EOM should be consistent with \eqref{EOM2}. Then with this consistence we can find the relationships between $\lambda_i$ and $I_j^{(nn')}$.

\begin{itemize}
\item {We first consider the mass term of the vector modes in action \eqref{effectiveAction}.  This term contains three parts, i.e., $I^{(nn)}_2\equiv\frac{1}{2}m_1^{(n)2},I^{(nn)}_4\equiv\frac{1}{2}m_2^{(n)2},I^{(nn)}_{11}\equiv\frac{1}{2}m_3^{(n)2} $. After comparing with the mass term in first line of \eqref{effequ11}, we can get $\lambda_1=-I^{(nn)}_2$, $\lambda_2=-I^{(nn)}_4$ and $\lambda_{3}=-I^{(nn)}_{11}$. By substituting in the definition of $\lambda_i$, we obtain three Schr\"{o}dinger-like equations
\begin{subequations}\label{vectorkk}
\begin{eqnarray}
  \left(\partial_{z,z}-V_{\text{eff1}}\right)W_1^{(n)}
  &=&m_1^{(n)2}\;W_1^{(n)},\label{Sch1}\\
   \left(\partial_{y,y}-V_{\text{eff2}}\right)W_1^{(n)}
  &=&m_2^{(n)2}\;W_1^{(n)},\label{Sch2}\\
   \left(\partial_{w,w}-V_{\text{eff3}}\right)W_1^{(n)}
  &=&m_3^{(n)2}\;W_1^{(n)},\label{Sch3}
\end{eqnarray}
\end{subequations}
with
\begin{subequations}\label{effectivePoten}
\begin{eqnarray}
V_{\text{eff1}}&=&\frac{3}{2}\;\partial_ {z,z}A+\frac{9}{4}\;\partial_{z}A\;\partial_{z}A,\label{effpoten1}\\
V_{\text{eff2}}&=&\frac{3}{2}\;\partial_{y,y} A+\frac{9}{4}\;\partial_{y}A\;\partial_{y}A,\label{effpoten2}\\
V_{\text{eff3}}&=&\frac{3}{2}\;\partial_{w,w} A+\frac{9}{4}\;\partial_{w}A\;\partial_{w}A.\label{effpoten3}
\end{eqnarray}
\end{subequations}
If the above Schr\"{o}dinger-like equations have solutions that satisfy the orthonormality condition \eqref{condition1}, there will be localized vector KK modes on the brane. These vector KK modes can obtain masses from the three extra dimensions $z, y, w$ respectively.  In brane with codimension-1 or codimension-2 \cite{Fu2015cfa, Fu2018erz}, the vector KK modes can also obtain masses from the extra dimensions, but the effective potentials are different from the \eqref{effectivePoten} .
}

\item {We now turn to the mass term of the three types of scalar KK modes $\phi^{(n)}$, $\varphi^{(n)}$ and $\chi^{(n)}$. From the effective action \eqref{effectiveAction} we can see that each of these scalars has two parts of masses. We denote these mass terms as $ I^{(nn)}_7\equiv m_{\phi1}^{(n)}, I^{(nn)}_{18}\equiv m_{\phi2}^{(n)}$, $I^{(nn)}_9\equiv m_{\varphi1}^{(n)}, I^{(nn)}_{19}\equiv m_{\varphi2}^{(n)}$ and $I^{(nn)}_{15}\equiv m_{\chi1}^{(n)}, I^{(nn)}_{17}\equiv m_{\chi2}^{(n)}$. Comparing these mass terms with those in \eqref{effequ22}$-$\eqref{effequ44}, and using the definition of $\lambda_i$, we get the following equations:
\begin{subequations}\label{scalarMasss}
\begin{eqnarray}
  \left(\partial_{y,y}-V_{\text{eff2}}\right)U_1^{(n)}
  &=&m_{\phi1}^{(n)2}\;U_1^{(n)},\label{Sch4}~~~~~
   \left(\partial_{w,w}-V_{\text{eff3}}\right)U_1^{(n)}
 =m_{\phi2}^{(n)2}\;U_1^{(n)},\label{Sch5}\\
  \left(\partial_{z,z}-V_{\text{eff1}}\right)U_2^{(n)}
  &=&m_{\varphi1}^{(n)2}\;U_2^{(n)},\label{Sch6}
~~~~~
   \left(\partial_{w,w}-V_{\text{eff3}}\right)U_2^{(n)}
  =m_{\varphi2}^{(n)2}\;U_2^{(n)},\label{Sch7}\\
  \left(\partial_{z,z}-V_{\text{eff1}}\right)U_3^{(n)}
  &=&m_{\chi1}^{(n)2}\;U_3^{(n)},\label{Sch8}~~~~~
   \left(\partial_{y,y}-V_{\text{eff2}}\right)U_3^{(n)}
  =m_{\chi2}^{(n)2}\;U_3^{(n)}.\label{Sch9}
\end{eqnarray}
\end{subequations}
The definitions of the effective potentials can be found in Eqs.~\eqref{effpoten1}-\eqref{effpoten3}.
If there are solutions for the equations \eqref{Sch5} satisfying the orthonormality condition\eqref{condition3}, the scalar $\phi^{(n)}$ can be localized on the brane, and can obtain masses from the extra dimensions $y$ and $w$. This can be easily understood, as the scalar $\phi^{(n)}$ KK mode is from $z$ component of the bulk field (see \eqref{kk2}). For the other two scalars $\varphi^{(n)}$ and $\chi^{(n)}$ we also have similar results.

The coincidence between the Schr\"odinger-like equations of the scalar modes with those of the vector modes indicates that there must be some relationships between $W_1^{(n)}$, $U_1^{(n)}, U_2^{(n)}$ and $U_3^{(n)}$}.

\item {These relationships can be found by studying the couplings between the vector and the  scalar KK modes. For simplicity, we will rename the coupling constants $I_{6}^{(nn)}$, $I_{8}^{(nn)}$ and $I_{12}^{(nn)}$ as $C_{1}^{(n)}$, $C_{2}^{(n)}$ and $C_{3}^{(n)}$, respectively. By comparing the two groups of EOM we  get the relations between $W_1^{(n)}$ and  $W_2^{(n)}, W_3^{(n)}, W_4^{(n)}$:
\begin{subequations}
\label{w1234}
\begin{eqnarray}I_{6}^{(nn)}&=&\lambda_{4}\equiv C_{1}^{(n)}=-\overline{\partial}_z\;\widetilde{U}^{(n)}_1/(2\overline{W}^{(n)}_1),\label{c21}\\
I_{6}^{(nn)}&=&\lambda_{7}\equiv C_{1}^{(n)}=2\widetilde{\partial}_z\;\overline{W}^{(n)}_1/{\widetilde{U}_1},\label{c22}
\\I_{8}^{(nn)}&=&\lambda_{5}\equiv C_{2}^{(n)}=-\overline{\partial}_y\;\widetilde{U}^{(n)}_2/(2\overline{W}^{(n)}_1),\label{c11}
\\I_{8}^{(nn)}&=&\lambda_{12}\equiv C_{2}^{(n)}=2\widetilde{\partial}_y\;\overline{W}^{(n)}_1/{\widetilde{U}_2},\label{c12}\\
I_{12}^{(nn)}&=&\lambda_{6}\equiv C_{3}^{(n)}=-\overline{\partial}_w\;\widetilde{U}^{(n)}_3/(2\overline{W}^{(n)}_1),\label{c31}\\
I_{12}^{(nn)}&=&\lambda_{17}\equiv C_{3}^{(n)}=2\widetilde{\partial}_w\;\overline{W}^{(n)}_1/{\widetilde{U}_3}.
 \label{c32}
 \end{eqnarray}
 \end{subequations}
With the above relations and the Schr\"{o}dinger-like equations \eqref{Sch1}$-$ \eqref{Sch3}, one can easily find the following results:
\begin{eqnarray}
C_{1}^{(n)2}&=&m_1^{(n)2},\\
C_{2}^{(n)2}&=&m_2^{(n)2},\\
C_{3}^{(n)2}&=&m_3^{(n)2},
\end{eqnarray}
which are similar to those obtained in models with codimension-1 and codimension-2, and are very important for the invariance of the  effective action. 

\item Finally, let us consider the couplings between the scalar KK modes.  We replace the coupling constants $I_{10}^{(nn)}$, $I_{14}^{(nn)}$, and $I_{16}^{(nn)}$ as $C_{4}^{(n)}$, $C_{5}^{(n)}$ and $C_{6}^{(n)}$ for convenience, and with the consistence of the two groups of EOM we find:
\begin{subequations}\label{wss}
\begin{eqnarray}I_{10}^{(nn)}&=&\lambda_{10}\equiv C_{4}^{(n)}=-2\overline{\partial}_y  (  \widetilde{\partial}_z\;\overline{U}^{(n)}_2)/{\overline{U}_1^{(n)}},\label{c41}\\I_{10}^{(nn)}&=&\lambda_{15}\equiv C_{4}^{(n)}=-2\overline{\partial}_z  (  \widetilde{\partial}_y\;\overline{U}^{(n)}_1)/{\overline{U}_2^{(n)}},\label{c42}\\I_{14}^{(nn)}&=&\lambda_{11}\equiv C_{5}^{(n)}=-2\overline{\partial}_w  (  \widetilde{\partial}_z\;\overline{U}^{(n)}_3)/{\overline{U}_1^{(n)}},\label{c51}\\I_{14}^{(nn)}&=&\lambda_{20}\equiv C_{5}^{(n)}=-2\overline{\partial}_z  (  \widetilde{\partial}_w\;\overline{U}^{(n)}_1)/{\overline{U}_3^{(n)}},\label{c52}\\I_{16}^{(nn)}&=&\lambda_{16}\equiv C_{6}^{(n)}=-2\overline{\partial}_w  (  \widetilde{\partial}_y\;\overline{U}^{(n)}_3)/{\overline{U}_2^{(n)}},\label{c61}\\I_{16}^{(nn)}&=&\lambda_{21}\equiv C_{6}^{(n)}=-2\overline{\partial}_y  (  \widetilde{\partial}_w\;\overline{U}^{(n)}_2)/{\overline{U}_3^{(n)}}.\label{c62}
\end{eqnarray}
\end{subequations}

By using the relationships between $W_1^{(n)}$ with $U_1^{(n)}, U_2^{(n)}, U_3^{(n)}$, i.e., Eqs.~\eqref{w1234} to the above equations,  one would finally obtain
\begin{eqnarray}
  C_{4}^{(n)2}=4m_{\phi1}^{(n)2}\;m_{\varphi1}^{(n)2}, \quad C_{2}^{(n)2}=m_{\phi1}^{(n)2},
\quad C_{1}^{(n)2}=m_{\varphi1}^{(n)2},\\
  C_{5}^{(n)2}=4m_{\phi2}^{(n)2}\;m_{\chi1}^{(n)2}, \quad C_{3}^{(n)2}=m_{\phi2}^{(n)2},
\quad C_{1}^{(n)2}=m_{\chi1}^{(n)2},\\
  C_{6}^{(n)2}=4m_{\varphi2}^{(n)2}\;m_{\chi2}^{(n)2}, \quad C_{2}^{(n)2}=m_{\chi2}^{(n)2},
\quad C_{3}^{(n)2}=m_{\varphi2}^{(n)2}.
\end{eqnarray}
Interestingly, we find that the masses of the scalar KK modes are related with the masses of the vector modes:
\begin{eqnarray}
m_{\phi}^{(n)2}=m_{2}^{(n)2}+m_{3}^{(n)2},~~
m_{\varphi}^{(n)2}=m_{1}^{(n)2}+m_{3}^{(n)2},~~
m_{\chi}^{(n)2}=m_{1}^{(n)2}+m_{2}^{(n)2}.
\end{eqnarray}
This is another crucial reason for the gauge invariance of the effective action.
}

\end{itemize}

Now with all above relationships between the coupling constants, we can simplified  the effective action \eqref{effectiveAction} as
\begin{eqnarray}\label{gaugeinvariantAction2}
  S&=&
     -\frac{1}{4}\sum_n\sum_{n}\int {d^{4}x} \sqrt{-\hat{g}}
     \;\bigg[\;\hat{Y}^{(n)}_{\mu_1\mu_2}\;\hat{Y}^{\mu_1\mu_2(n')}\nonumber\\
     &&+2\big((\partial_{\mu}\phi^{(n)}-\frac{1}{2}m_1^{(n)}\;\hat{X}_{\mu}^{(n)})^2
    +(\partial_{\mu}\varphi^{(n)}-\frac{1}{2}m_2^{(n)}\;\hat{X}_{\mu}^{(n)})^2
    +(\partial_{\mu}\chi^{(n)}-\frac{1}{2}m_3^{(n)}\;\hat{X}_{\mu}^{(n)})^2\big)\nonumber\\
    &&+\big(m_2^{(n)}\phi^{(n)}-m_1^{(n)}\varphi^{(n)}\big)^2
    +\big(m_2^{(n)}\chi^{(n)}-m_3^{(n)}\varphi^{(n)}\big)^2
    +\big(m_3^{(n)}\phi^{(n)}-m_1^{(n)}\chi^{(n)}\big)^2
    \bigg],
\end{eqnarray}
which is invariant under the following gauge transform
\begin{eqnarray}
   &&\hat{X}_{\mu}^{(n)}\rightarrow \hat{X}_{\mu}^{(n)}+\partial_{\mu}\rho^{(n)},\nonumber \\
   &&\phi^{(n)}\rightarrow \phi^{(n)}+\frac{1}{2}m_1^{(n)}\rho^{(n)}, ~~
    \varphi^{(n)}\rightarrow \varphi^{(n)}+\frac{1}{2}m_2^{(n)}\rho^{(n)},~~
    \chi^{(n)}\rightarrow \chi^{(n)}+\frac{1}{2}m_3^{(n)}\rho^{(n)}.
  \end{eqnarray}

\section{Generalization in a brane model with an arbitrary number of extra dimensions }

In this section, we consider a model with $d$ extra dimensions ($d=1, 2...$). For this case the KK decomposition can also be written as \eqref{kkdecomposition} with $a=-d/2$, and there will be  $d$ kinds of the scalar KK modes, for which we can labeled as $\phi^{(n)}_i~(i=1,2...d)$, as there are $d$ extra dimensional components  $X_i$ for the bulk vector field. Then we can repeat the calculation as the case for a brane with three extra dimensions.

Referring to the effective action \eqref{effectiveacion1}, for this case, we can classify the coupling constants into four types. Two types are the self-couplings of the vector and scalar KK modes, they are related to the masses of the KK modes. For the vector ones they can obtain masses from all the extra dimensions, so we can mark the coupling constant as $I_{\text{vec}}^{(n)}=\sum_{i} \frac{1}{2} m_i^{(n)2}$ . For the scalar ones, they only can obtain masses from $(d-1)$ extra dimensions, because they come from the $\sum \sum_{i,j} Y_{ij}Y^{ij}~(j\neq i, j=1,2...d)$ terms of the bulk action. So that for $d=1$, the scalar modes are all massless, and for $d=2$,  $Y_{12}Y^{12}$ is the only term that contains the scalar masses. Here we can introduce the mass parameters for the scalar mode $\phi^{(n)}_i$ as $I_{i}^{(n)}=\sum \widetilde{m}_{ik}^{(n)2}~(k\neq i, k=1,2...d)$. The third type couplings (denoted as $\bar{I}_i^{(n)}$) are those between the vector and scalar KK modes, and the fourth type (noted by $\bar{\bar{I}}_{ij}^{(n)}$) are couplings between different types of scalars $(\phi_i^{(n)}$ and  $\phi_j^{(n)}$ $( j\neq i)$.

Then we can derive the EOM from the effective action:
\begin{subequations}\label{EOMd}
\begin{eqnarray}
&&\frac{1}{\sqrt{-\hat{g}}}\;\partial_{\mu_1}
 \big (\sqrt{-\hat{g}}\;\hat{Y}^{\mu_1\mu_2(n)}\big)
 +\sum_{i} \frac{1}{2} m_i^{(n)2}\;\hat{X}^{\mu_2(n)}
 -\sum_{i}\bar{I}_i^{(n)}\;\partial^{\mu_2}\phi^{(n)}_i=0, \label{effequd1}\\
&&\frac{1}{\sqrt{-\hat{g}}}\;\partial_{\mu_1}
  \big(\sqrt{-\hat{g}}\;\partial^{\mu_1}\phi^{(n)}_i
  -\bar{I}_i^{(n)}\;\sqrt{-\hat{g}}\;\hat{X}^{\mu_1(n)}\big)
  +\sum_{k\neq i} \widetilde{m}_{ik}^{(n)2}\;\phi^{(n)}_i
  -\sum_{j\neq i}\bar{\bar{I}}_{ij}^{(n)}\;\phi^{(n)}_j=0,\label{effequd2}
\end{eqnarray}
\end{subequations}
Because there are $d$ kinds of scalar modes, there are $d+1$ sub-equations for the above EOM.  On the other hand, if we substitute the KK decomposition into the EOM for the bulk vector field, we can get another group of EOM for the KK modes, which should be consistent with Eq. \eqref{EOMd}. By comparing these two groups of equations, we can get the dynamical equations for the vector and scalar KK modes, and the relationships between the coupling constants.

First, for the massive vector KK modes, with an orthonormality condition
$\int dz_1\;dz_1\dots dz_d \;W_1^{(n)}W_1^{(n')}=\delta_{nn'}$, they satisfy the following Schr\"{o}dinger-like equation
\begin{eqnarray}\label{schod}
  \left(\partial_{z_i,z_i}-V_{i}\right)W_1^{(n)}
  &=&m_i^{(n)2}\;W_1^{(n)},\label{Sch1}
\end{eqnarray}
with
\begin{eqnarray}\label{vi}
V_{i}=\frac{d}{2}\;\partial_ {z_i,z_i}A+\frac{d^2}{4}\;\partial_{z_i}A\;\partial_{z_i}A.
\end{eqnarray}
It can be easily checked that for $d=1, 2$ the results are the same with that in  Refs. \cite{Fu2015cfa, Fu:2016vaj}. And for $d=3$ it is just Eq. \eqref{vectorkk}. Second for the vector-scalar couplings, it can be got that
$\bar{I}_i^{(n)}=-\overline{\partial}_{z_i}\widetilde{U}^{(n)}_{i}/(2\overline{W}^{(n)}_1)=
2\widetilde{\partial}_{z_i}\overline{W}^{(n)}_1/{\widetilde{U}_{i}}$, where we have defined  $\overline{U}^{(n)}_{i}\equiv \text{e}^{-\frac{d}{2}A} U^{(n)}_{i}$,
$\widetilde{U}^{(n)}_{i}\equiv\text{e}^{\frac{d}{2}A}U^{(n)}_{i}$ and
$\overline{\partial}_{z_i}\equiv\text{e}^{-dA}\partial_{z_i}$,
$\widetilde{\partial}_{z_i}\equiv\text{e}^{dA}\partial_{z_i}$. For $d=3$, they are Eqs. \eqref{c21}$\thicksim$\eqref{c32}. Thus, the Schr\"{o}dinger-like equations \eqref{schod} just tells us that $\bar{I}_i^{(n)}=m_i^{(n)}$. At the same time, for the scalar-scalar couplings, it is found that
$\bar{\bar{I}}_{ij}^{(n)}=\overline{\partial}_{z_j}  ( \widetilde{\partial}_{z_i} \overline{U}^{(n)}_j)/{\overline{U}_i^{(n)}}
=\overline{\partial}_{z_i}  (  \widetilde{\partial}_{z_j} \overline{U}^{(n)}_i)/{\overline{U}_j^{(n)}}$, which implies that $\bar{\bar{I}}_{ij}^{(n)}=4m_{i}^{(n)}m_{j}^{(n)}$ and $\widetilde{m}_{ik}^{(n)}=m_{k}^{(n)}$. These results also can be checked for $d=1,2,3$ that we have got.

These relationships between the couplings constants finally lead to an effective action as
 \begin{eqnarray}\label{gaugeinvariantAction2}
  S&=&
     -\frac{1}{4}\sum_n\sum_{n}\int {d^{4}x} \sqrt{-\hat{g}}
     \;\bigg[\;\hat{Y}^{(n)}_{\mu_1\mu_2}\;\hat{Y}^{\mu_1\mu_2(n')}+
     \sum_i\sum_j\bigg(2(\partial_{\mu}\phi^{(n)}_i-\frac{1}{2}m_i^{(n)}\;\hat{X}_{\mu}^{(n)})^2
    +(m_i^{(n)}\phi^{(n)}_j-m_j^{(n)}\phi^{(n)}_i)^2\bigg)
    \bigg],
\end{eqnarray}
which is gauge invariant under gauge  transform $
\hat{X}_{\mu}^{(n)}\rightarrow \hat{X}_{\mu}^{(n)}+\partial_{\mu}\rho^{(n)},
\phi^{(n)}_i\rightarrow \phi^{(n)}_i+\frac{1}{2}m_i^{(n)}\rho^{(n)} $. Moreover, considering the effective potential \eqref{vi} it needs the solutions of the brane model with $d$ extra dimensions to calculate how many masses the KK modes will obtain.

\section{Summary and discussions}
\label{conclusion}

In this work, we investigated the KK modes of a bulk massless $U(1)$ gauge field for a brane model with $d$ extra dimensions.  Our results are summarized as follows:
\begin{itemize}
\item We first considered the case with three extra dimensions. Using a general KK decomposition \eqref{kkdecomposition} without choosing any gauge for the bulk $U(1)$ gauge field,  we found four types of KK modes, i.e., one vector and three types of scalar. In addition to their self-interactions, these modes also coupled mutually, as can be seen in \eqref{effectiveAction}, thus there are many coupling constants in the effective action, which should be not independent if the effective action is invariant; Then by comparing the EOM derived from two approaches, we found that all the KK modes are constrained by some Schr\"{o}dinger-like equations, for which the corresponding effective potentials depend on the warp factor of the background geometry. It was also found some  precise relationships between the coupling constants in the effective action \eqref{effectiveAction}, which also are the relationships between $W_1^{(n)}, U_1^{(n)}, U_2^{(n)}$ and $U_3^{(n)}$. With these relationships we simplified the effective action for the $U(1)$ gauge field and proved that it is gauge invariant. This action implies that  the vector KK modes can obtain masses from the three extra dimensions, but the scalar ones only can obtain masses from two of the three extra dimensions. The masses of the scalars are related with the vectors;

\item Then we generalized the above result to a brane model with codimension-$d$. We have classified the coupling constants into four types, and finally found the relationships between them, with which we got a gauge invariant effective action. The Schr\"{o}dinger-like equations satisfied by the vector KK modes were derived, and the corresponding effective potentials depend on $d$.

\end{itemize}

Although in this work, we did not consider the solutions of the brane world models, and mainly focused on the gauge invariance of the effective actions of the vector KK modes, we can simply discuss the localization of the KK modes in some brane world models. In Refs. \cite{Freitas:2020mxr, Freitas:2020vcf} the authors found that when consider the effects of the Einstein equations on the bulk vector field, the free vector KK modes cannot be localized on the brane unless introducing some special mechanisms.  But if assuming that the field is a perturbation and ignoring the back-reaction effect on the background geometry, some of the present authors have realized the localization of vector zero mode in a model with a finite extra dimension \cite{LocalizationFuPRD2011, PRDefromedLiu} . Since the present work does not assume the specific solution of the metric, the results of gauge invariance obtained here can be applied to the model of Refs \cite{LocalizationFuPRD2011, PRDefromedLiu}, after taking $d=1$. For models with more extra dimensions, there is no solutions available for checking the localization of the vector zero mode. And this issue deserves another independent work.  And we know that it is not easy to localize the massive vector KK modes. But some of the present authors have found, in Ref. \cite{LocalizationFuPRD2011}, that a coupling between the bulk vector with a dilation field would lead to a P\"{o}schl-Teller (PT) potential for the vector KK modes. This kind of potential approaches to a positive constant at the infinite, which indicates a possibility of localized massive modes. It is interesting to extend the present work to a model with vector-dilaton coupling. We will leave this topic to our future work.

\chapter{\textbf{Acknowledgments}}

This work was supported by the National Natural Science Foundation of China (Grants No. 11405121, No. 11605127, 11847211), and the Fundamental
Research Funds for the Central Universities (Grant No. xzy012019061, Grant No. xzy012019052).




\end{document}